\documentclass[letterpaper, 10 pt, conference]{ieeeconf}
\IEEEoverridecommandlockouts
\usepackage{amsmath,amssymb,amsfonts}
\usepackage{algorithmic}
\usepackage{graphicx}
\usepackage{textcomp}
\usepackage{xcolor}
\usepackage{mathtools}
\usepackage{ntheorem}
\usepackage{epstopdf}
\usepackage{caption}
\usepackage{subcaption}
\usepackage{setspace}
\usepackage{subcaption}
\usepackage{dsfont}
\usepackage{cleveref}
\usepackage{tabularx,ragged2e,booktabs,caption}
\makeatletter
\newcommand*{\rom}[1]{\expandafter\@slowromancap\romannumeral #1@}
\makeatletter
\renewcommand*\env@matrix[1][*\c@MaxMatrixCols c]{%
  \hskip -\arraycolsep
  \let\@ifnextchar\new@ifnextchar
  \array{#1}}
\makeatother

\newcommand{\Real}{{\mathds R}} 
\newcommand{\Nat}{{\mathds N}} 

{}
{}
\newtheorem{proposition}{Proposition}{}
{}
\newtheorem{theorem}{Theorem}{}
\newtheorem{remark}{Remark}{}
\newtheorem{lemma}{Lemma}{}
{}

\def\BibTeX{{\rm B\kern-.05em{\sc i\kern-.025em b}\kern-.08em
    T\kern-.1667em\lower.7ex\hbox{E}\kern-.125emX}}
\begin{document}

\title{Gaussian Mechanisms Against Statistical Inference: Synthesis Tools}

\author{Haleh Hayati, Carlos Murguia, Nathan van de Wouw%
\thanks{Haleh Hayati, Carlos Murguia, and Nathan van de Wouw are with the Department of Mechanical Engineering, Dynamics and Control Group, Eindhoven University of Technology, The Netherlands. Emails: \& h.hayati@tue.nl, \& c.g.murguia@tue.nl, \& n.v.d.wouw@tue.nl.}
}
\maketitle

\begin{abstract}
In this manuscript, we provide a set of tools (in terms of semidefinite programs) to synthesize Gaussian mechanisms to maximize privacy of databases. Information about the database is disclosed through queries requested by (potentially) adversarial users. We aim to keep part of the database private (private sensitive information); however, disclosed data could be used to estimate private information. To avoid an accurate estimation by the adversaries, we pass the requested data through distorting (privacy-preserving) mechanisms before transmission and send the distorted data to the user. These mechanisms consist of a coordinate transformation and an additive dependent Gaussian vector. We formulate the synthesis of distorting mechanisms in terms of semidefinite programs in which we seek to minimize the mutual information (our privacy metric) between private data and the disclosed distorted data given a desired distortion level -- how different actual and distorted data are allowed to be.\linebreak 
\end{abstract}


\section{INTRODUCTION}
Scientific and technological advancements in today's hyperconnected world have resulted in a massive amount of user data collected and processed by hundreds of companies over public networks. Companies use this data to provide personalized services and target advertising. However, these technologies have come with the cost of a significant loss of privacy in society. Depending on their resources, adversaries can infer sensitive (private) information about databases from public data available on the internet and/or unsecured networks/servers. A motivating example of this privacy loss is the potential use of data from smart electrical meters by criminals, advertising agencies, and governments for monitoring the activities of occupants \cite{Poor1,Poor2}.  Another example is the privacy loss caused by information sharing in distributed control systems and cloud computing \cite{Huang:2014:CDP:2566468.2566474}.
This is why researchers from several fields (e.g., computer science, information theory, and control theory) have been drawn to the broad research subject of privacy and security of Cyber-Physical Systems (CPSs), \cite{Farokhi1}-\nocite{Farokhi2}\nocite{Pappas}\nocite{Jerome1}\nocite{Takashi_1}\nocite{Takashi_3}\nocite{chaper_privacy_chaos}\cite{Carlos_Iman1}.

We consider the following setting. Users request data, say $Y$, from public databases through a set of queries. There is sensitive (private) information in the database, say $S$, that we aim to keep private. However, due to the statistical dependence between $Y$ and $S$, adversarial users could use disclosed data to infer private information. To avoid accurate inference, we pass the requested data through distorting (privacy-preserving) mechanisms before transmission and send the distorted data to the user. These mechanisms consist of a linear coordinate transformation, $G$, and an additive dependent Gaussian vector, $V$; Therefore, the distorted disclosed data, $Z$, is given by $Z = GY+V$, $V \sim \mathcal{N}[\mathbf{0},\Sigma^V]$, for some covariance matrix $\Sigma^V$. We formulate the synthesis of the distorting mechanisms (the pair $(G,\Sigma^V)$) in terms of semidefinite programs in which we seek to conceal (as much as possible) the private parts of the database.
Here we follow an information-theoretic approach to privacy. As \emph{privacy metric}, we use the \emph{mutual information}, $I[S;Z]$, between private data, $S$, and the disclosed distorted data, $Z$. Mutual information is a measure of the statistical dependence between $S$ and $Z$ \cite{Cover}. We cast the design of $(G,\Sigma^V)$ as a semidefinite program where we minimize $I[S;Z]$. Note, however, that it is not desired to overly distort the original data. We might change original data excessively to maximize privacy. Therefore, when synthesizing $(G,\Sigma^V)$ (by minimizing $I[S;Z]$), we have to consider the trade-off between \emph{privacy} and \emph{distortion}. As \emph{distortion metric}, we use the \emph{weighted mean squared error} between the original data, $Y$, and its distorted version, $Z$. Therefore, the overall optimization problem amounts to minimizing $I[S;Z]$, subject to an upper bound on the mean squared error (how different $Y$ and $Z$ are allowed to be on average), using $(G,\Sigma^V)$ as optimization variables.

Using additive random vectors is common to promote privacy by avoiding an accurate estimation of private data. Differential privacy is popular approach in the context of privacy of databases \cite{Jerome1,Dwork}. In differential privacy, because it provides certain privacy guarantees, Laplace noise is commonly employed \cite{Dwork2}, without tackling the problem of finding the maximum level of privacy for a given allowed distortion level. Generally, the additive noise distribution that yields maximum privacy is determined by the particular privacy and distortion metrics being considered, as well as the system configuration \cite{Topcu}-\nocite{SORIA}\nocite{Geng}\cite{Dullerud}. There are results addressing this subject from an information-theoretic perspective, in which privacy is quantified using information metrics -- e.g., mutual information, entropy, Kullback-Leibler divergence, and Fisher information \cite{Poor1,Farokhi1,Farokhi2,FAROKHI3,Fawaz,Fawaz2,Lalita}.\\
If the data to be kept private follows continuous probability distributions, the problem of finding the optimal additive noise to maximize privacy is difficult to solve. This issue has been addressed by assuming the data to be kept private is deterministic \cite{Farokhi1,SORIA,Geng}. However, in the context of databases, unavoidable correlation among features results in statistical dependence and, therefore, deterministic tools often lead to conservative results. Recently, the authors in \cite{Carlos_Iman1,murguia2020privacy,murguia2021privacy} have proposed a framework for synthesizing optimal transition probabilities to maximize privacy for a class of Cyber-Physical Systems (CPSs) characterized by discrete multivariate probability distributions. Although this framework is fairly general and leads to distorting mechanisms following arbitrary distributions, the computational complexity induced by exploring all the possible transition probabilities from private to disclosed distorted data is high and grows exponentially with the alphabet of the private data. In \cite{Farokhi1}, the authors neglect quantization and work directly with dynamical systems driven by continuous (Gaussian) disturbances. They prove that in the case of unconstrained additive noise, the optimal noise distribution minimizing Fisher information is Gaussian. When employing mutual information as a privacy metric, the same statement was also proved in \cite{Cedric}. \\
Motivated by these results, we propose a general class of Gaussian mechanisms and seek the optimal parameterization. In \cite{cdc2021arxiv,hayati2021finite}, we present an optimization-based framework for synthesizing optimal privacy-preserving Gaussian mechanisms for a class of stochastic dynamical systems. Our first contribution in this paper is to extend the optimization-based framework for synthesizing Gaussian mechanisms in the context of databases. Furthermore, we prove that we can allow for prior distributions (the joint probability distributions between private and disclosed data) that are not necessarily Gaussian, as long as they are log-concave \cite{prekopa1980logarithmic}. The latter generalizes further the class of databases our framework can deal with.\\
The structure of the paper is organized as follows. The problem formulation, including adversarial capabilities and privacy and distortion metrics, is presented in Section \rom{2}. The solution to the Problem $1$ formulated in Section \rom{2} is developed in Section \rom{3}. Our simulation results are presented in Section \rom{4}, followed by concluding remarks in Section \rom{5}.\\
\textbf{Notation:} The notation $X \sim \mathcal{N}[\mu,\Sigma^X]$ stands for a normally distributed random vector $X \in \Real^{n}$ with mean $E[X] = \mu \in \Real^{n}$ and covariance matrix $E[(X-\mu)(X-\mu)^T] = \Sigma^X \in \Real^{n \times n}$, where $E[a]$ denotes the expected value of the random vector $a$. The $n \times n$ identity matrix is denoted by $I_n$ or simply $I$ if $n$ is clear from the context. Similarly, $n \times m$ matrices composed of only zeros are denoted by $\mathbf{0}_{n \times m}$ or simply $\mathbf{0}$ when their dimensions are clear. For positive definite (semidefinite) matrices, we use the notation $P>0$ ($P \geq 0$). The operators $\log[\cdot]$, $\det[\cdot]$, and $\text{tr}[\cdot]$ stand for logarithm base two, determinant, and trace, respectively.

\section{PROBLEM FORMULATION}
\begin{figure*}
  \centering
  \includegraphics[width=0.7\textwidth]{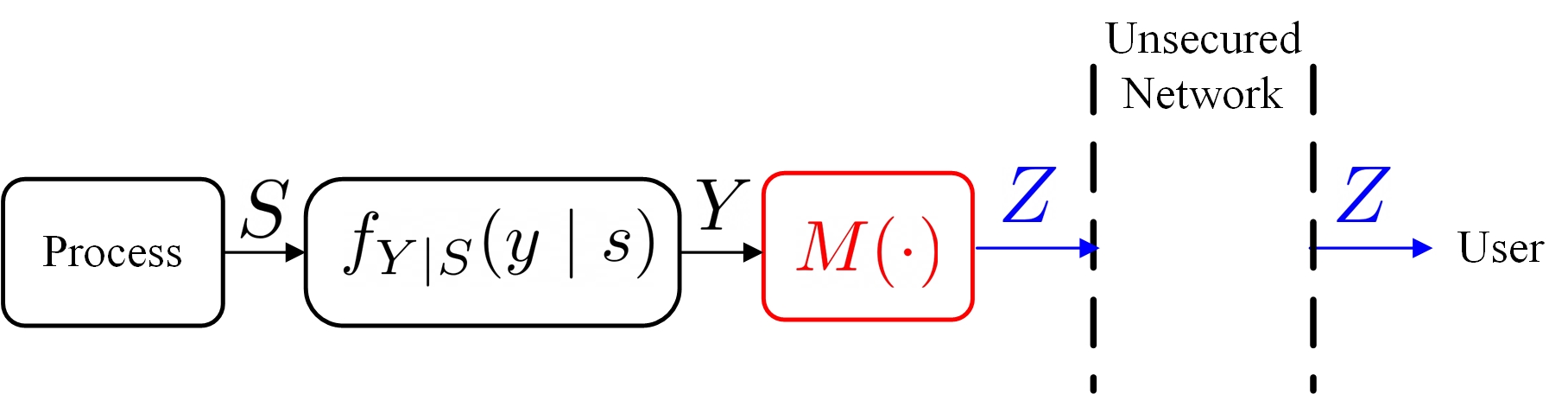}
    \caption{System configuration.}
    \label{fig1}
\end{figure*}

Let $S$ be part of a database $D$ that must be kept private. We model $S$ as a multivariate Gaussian distribution, $S \sim \mathcal{N}[\mu_S,\Sigma^S]$, with mean $\mu_S \in \mathbb{R}^{n_s}$ and covariance $\Sigma^S \in \mathbb{R}^{n_s \times n_s}$, $\Sigma^S >0$, $n_{s} \in \Nat$. Information about $S$ is obtained through queries of the form $Y=q(S)$, for some (stochastic or deterministic) affine mapping $q: \mathbb{R}^{n_{s}} \rightarrow \mathbb{R}^{n_{y}}$ characterized by transition probabilities $f_{Y \mid S}(y \mid s)$. We model $Y$ as a multivariate Gaussian distribution, $Y \sim \mathcal{N}[\mu_Y,\Sigma^Y]$, with mean $\mu_Y \in \mathbb{R}^{n_y}$ and covariance $\Sigma^Y \in \mathbb{R}^{n_y \times n_y}$, $\Sigma^Y>0$, $n_{y} \in \Nat$. To capture statistical dependence between private and disclosed data, we let $S$ and $Y$ be jointly distributed as $\begin{psmallmatrix} Y \\ S  \end{psmallmatrix} \sim \mathcal{N}\left[\mu^{Y,S},\Sigma^{Y,S}\right]$ with mean $\mu^{Y,S} \in \mathbb{R}^{(n_s +n_y)}$, joint covariance $\Sigma^{Y,S}\in \mathbb{R}^{(n_s +n_y) \times (n_s +n_y)}$, $\Sigma^{Y,S}>0$, and cross-covariance $\Sigma^{YS}\in \mathbb{R}^{n_y \times n_s}$, $\Sigma^{YS}>0$.\\
We aim to prevent adversaries from accurately estimating the private data $S$. To this end, we randomize $Y$ before disclosure and send the corrupted data to the user instead. We randomize $Y$ through a mapping $M(\cdot)$ of the form:
\begin{equation}
    Z=M(Y):=GY+V,\label{eq2}
\end{equation}
for some linear transformation $G \in {\mathbb{R}^{{n_y} \times {n_y}}}$ and dependent multivariate Gaussian vector $V \sim \mathcal{N}[\mathbf{0},\Sigma^V]$ with covariance $\Sigma^V \in \mathbb{R}^{n_y \times n_y}$, $\Sigma^V>0$. The randomized vector $Z$ is disclosed to the user, see Figure \ref{fig1}. We seek to synthesize distortion variables $(G,\Sigma^V)$ to make inference of private data, $S$, as challenging as possible from the disclosed data, $Z$.
\subsection{Adversarial Capabilities}
We consider worst-case adversaries that send queries to the server or eavesdrop data at the communication channel. They have access to distorted data $Z$, and have prior knowledge of the statistical properties of the database, i.e., the adversary knows $(\mu_S,\Sigma^S,\mu_Y,\Sigma^Y,\Sigma^{Y,S})$. We assume the adversary uses a linear Minimum Mean Square Error (MMSE) estimator to reconstruct $S$ from $Z$ and prior knowledge. The MMSE estimator offers the best estimation performance among all unbiased estimators for jointly Gaussian vectors \cite{huemer2017component}. In practice, actual adversaries are unlikely to have all the capabilities we assume here. However, if we maximize privacy against these worst-case adversaries, we ensure that adversaries with fewer capabilities perform even worse (or equal at most).

\subsection{Metrics and Problem Formulation}
The aim of our privacy scheme is to make inference of $S$ from the distorted disclosed data $Z$ as challenging as possible without distorting $Y$ excessively. That is, we do not want to make $Y$ and $Z$ overly different because of the utility of the query output $Y$ for users. Hence, when designing the distorting variables $(G,\Sigma^V)$, we need to consider the \emph{trade-off between privacy and distortion}. As distortion metric, we use the weighted mean squared error between the original and distorted data, i.e., $E[||W(Z - Y)||^2]$ for a given weighting matrix $W$ of appropriate dimension. Matrix $W$ is used to fine-tune the desired distortion at different channels and/or to model different applications of the distorted data. As a privacy metric, we use the mutual information between private and disclosed data, i.e., $I[S;Z]$, which is a measure of the statistical dependence between $S$ and $Z$ \cite{Cover}. Hence, we aim at minimizing $I[S;Z]$ subject to the second moment constraint $E[||W(Z - Y)||^2] \leq \epsilon$, for a desired maximum distortion level ${\epsilon} \in {\mathbb{R}^+}$, by designing $G$ and ${\Sigma^V}$ of the distorting mechanisms \eqref{eq2}. In what follows, we present the optimization problem we seek to solve.\\

\textbf{Problem 1}: Given the private $S$, disclosed $Y$, desired maximum distortion level ${\epsilon} \in {\mathbb{R}^+}$, weighting matrix ${W} \in {\mathbb{R}^{{n_y} \times {n_y}}}$, and the distorting mechanism \eqref{eq2}, find the distorting variables, $G$ and ${\Sigma^V}$, as the solution to:
\begin{equation}\label{problem1}
    \left\{\begin{aligned}
&\min _{G,\Sigma^V}  \hspace{1mm} I[S;Z], \\
&\text { s.t. }  E[||W(Z - Y)||^2] \leq \epsilon, \\[1.5mm]
&\hspace{6.5mm}\text{$V$ and $Y$ independent.}
\end{aligned}\right.
\end{equation}
\section{SOLUTION TO PROBLEM 1}
To solve Problem 1, we first need to write the cost function and constraint in terms of the design variables $G$ and $\Sigma^V$.
\subsection{Cost Function: Formulation and Convexity}
The mutual information $I\left[S;Z \right]$ can be written in terms of differential entropies as $I[S;Z] = h[S] + h[Z] - h[S,Z]$ \cite{Cover}. Moreover, differential entropies of Gaussian random vectors are fully characterized by their covariance matrices, $h[S] = \frac{1}{2}\log \det \left( \Sigma^S \right) + \frac{n_s}{2} + \frac{n_s}{2}\log(2\pi)$, (see Theorem $8.4.1$ in \cite{Cover}). So, to characterize $I\left[ S;Z \right]$ in terms of $G$ and $\Sigma^V$, we need to write the covariance matrices of $S$ and $Z$ and their joint covariance in terms of design variables, $(G,\Sigma^V)$.\\
In the following lemma, we give a closed-form expression of the joint density of $Z$ and $S$.

\begin{lemma}\label{stackedDist}
\[ \begin{psmallmatrix} Z \\ S  \end{psmallmatrix} \sim \mathcal{N}\left[\mu^{Z,S},\Sigma^{Z,S}\right],\] with mean $\mu^{Z,S} \in \mathbb{R}^{(n_s + n_y)}$ and joint covariance matrix $\Sigma^{Z,S} \in \mathbb{R}^{(n_s + n_y) \times (n_s + n_y)}$, $\Sigma^{Z,S}>0$, given by\emph{:}
\begin{align}
\mu^{Z,S}&=\left[\begin{array}{c}
G\mu_Y \\
\mu_S
\end{array}\right] ,\label{muZS}\\
\Sigma^{Z,S}&=\left[\begin{array}{cc}
\Sigma^{Z} & \Sigma^{ZS} \\
{\Sigma^{ZS}}^{\top} & \Sigma^{S}
\end{array}\right] ,\label{SigmaZS} \\
\Sigma^{Z} &= G \Sigma^Y G^\top + \Sigma^V, \label{SigmaZ}\\
\Sigma^{ZS}&=G \Sigma^{YS},\label{CovZS}
\end{align}
where $\Sigma^{ZS}$ denotes the cross-covariance between $Z$ and $S$.
\end{lemma}
\emph{\textbf{Proof}}: Because $Y$ and $V$ are independent, $G$ is deterministic, and $E[V]=\mathbf{0}$, we have $Z=GY+V \sim \mathcal{N} [ {\mu_Z , \Sigma^Z} ]$, with mean $\mu_Z=G\mu_Y$ and covariance $\Sigma^Z$ as in \eqref{SigmaZ}, see \cite{Ross} for details. To simplify notation, we introduce the stacked vector ${\Theta} := {( {{{Z}^\top},{{S}^\top}} )^\top}$. It follows that $E[ {{\Theta }} ] = \mu^{Z,S}$, with $\mu^{Z,S}$ as in \eqref{muZS}. By definition of (joint) covariance and cross-covariance \cite{Ross}, we can write:
\begin{equation}\label{covdefinition}
    \begin{aligned}
\Sigma^{Z,S} &= E\left[{\Theta}{\Theta}^{\top}\right]
-E\left[{\Theta}\right] E\left[{\Theta}^{\top}\right]\\
&=\left[\begin{array}{ll}
E\left[Z Z^{\top}\right]-\mu_{Z} \mu_{Z}^{\top} & E\left[Z S^{\top}\right]-\mu_{Z} \mu_{S}^{\top} \\[1mm]
E\left[S Z^{\top}\right]-\mu_{S} \mu_{Z}^{\top} & E\left[S S^{\top}\right]-\mu_{S} \mu_{S}^{\top}
\end{array}\right],
\end{aligned}
\end{equation}
and  $\Sigma^{Z}=E\left[Z Z^{\top}\right]-\mu_{Z} \mu_{Z}^{\top}$, $\Sigma^{S}=E\left[S S^{\top}\right]-\mu_{S} \mu_{S}^{\top}$, and $\Sigma^{ZS}=E\left[Z S^{\top}\right]-\mu_{Z} \mu_{S}^{\top}$. Therefore, \eqref{covdefinition} can be written in terms of these covariances as in \eqref{SigmaZS}. The cross-covariance $\Sigma^{ZS}$ can be expanded as follows:
\begin{subequations}
\begin{align}
\Sigma^{ZS}&=E\left[(G Y+V) S^{\top}\right]- \left(G\mu_Y{\mu_S^{\top}}\right)\\
&=G\left(E\left[Y S^{\top}\right]-\left(\mu_Y{\mu_S^{\top}}\right)\right)=G \Sigma^{YS},\label{eqcrosscov2}
\end{align}
\end{subequations}
where \eqref{eqcrosscov2} follows from the independence between $V$ and $S$ (thus $E[VS^\top]=0$), and $\Sigma^{YS}$ denotes the cross-covariance between $Y$ and $S$.\\
So far, we have written the mean and covariance of ${\Theta}$ as in \eqref{muZS} and \eqref{SigmaZS}, respectively. It remains to prove that ${\Theta}$ follows the Gaussian distribution $\mathcal{N}\left[\mu^{Z,S},\Sigma^{Z,S}\right]$. The joint probability distribution of two Gaussian vectors is Gaussian if and only if their joint covariance matrix is not degenerate (i.e., $\Sigma^{Z,S} > 0$), see \cite{Ross} for details. Necessary and sufficient conditions for $\Sigma^{Z,S}>0$ are that $\Sigma^S$ is positive definite (which is true by assumption) and the Schur complement of block $\Sigma^S$ of $\Sigma^{Z,S}$ in \eqref{SigmaZS}, denoted as $\Sigma ^{Z,S} / \Sigma^S$, is positive definite (\cite{zhang2006schur}, Theorem 1.12). This Schur complement is given by:
\begin{subequations}
\begin{align}
\Sigma^{Z,S} / \Sigma^S &=
G \Sigma^Y G^\top + \Sigma^V -G \Sigma^{YS} {\Sigma^S}^{-1} {\Sigma^{YS}}^\top G^\top \\
&= G(\Sigma^Y - \Sigma^{YS} {\Sigma^S}^{-1} {\Sigma^{YS}}^\top)G^\top + \Sigma^V.
\end{align}\label{SigmaZSschur}\end{subequations}
By definition, the joint covariance of $Y$ and $S$, $\Sigma^{Y,S}$, is positive definite (because they are jointly Gaussian); hence, its Schur complement $\Sigma ^{Y,S} / \Sigma^S$ is also positive definite \cite{zhang2006schur}. Matrix $\Sigma ^{Y,S} / \Sigma^S$ is given by
\begin{equation}
\Sigma^{Y,S} / \Sigma^S = \Sigma^Y - \Sigma^{YS} {\Sigma^S}^{-1} {\Sigma^{YS}}^\top>0.\label{ineqSigmaYS}
\end{equation}
Therefore, from \eqref{ineqSigmaYS}, the fact that $\Sigma^V>0$ (by construction), and \eqref{SigmaZSschur}, we can conclude that $\Sigma^{Z,S} / \Sigma^S$ in \eqref{SigmaZSschur} is positive definite and hence, $\Sigma^{Z,S}>0$.
 \hfill $\blacksquare$\\
In Lemma \ref{stackedDist}, we have written $(\Sigma^Z,\Sigma^S,\Sigma^{S,Z})$ in terms of the design variables $G$ and $\Sigma^V$. Before we write the cost in \eqref{problem1} in terms of these matrices, we notice that $\Sigma^V$ only appears in the expression for $\Sigma^Z$ in \eqref{SigmaZ}. Moreover, given $(G,\Sigma^Z)$, matrix $\Sigma^V$ is fully determined and vice versa. That is, $(G,\Sigma^V) \rightarrow (G,\Sigma^Z)$ is an invertible transformation, and therefore, we can pose both cost and constraints in terms of either $\Sigma^V$ or $\Sigma^Z$. By casting the problem in terms of $\Sigma^Z$, we can establish a linear distortion constraint and a convex cost function. Hereafter, the problem is posed in terms of $(G,\Sigma^Z)$. Then, we extract the optimal $\Sigma^V$ from the optimal $(G,\Sigma^Z)$ using \eqref{SigmaZ}. The first constraint that we need to enforce is that the extracted $\Sigma^V$ is always positive definite (as it is a covariance matrix). From \eqref{SigmaZ}, it is easy to verify that $\Sigma^V>0$ if and only if $\Sigma^{Z} - G \Sigma^Y G^\top  > 0$. Using standard Schur complement properties \cite{zhang2006schur}, the latter nonlinear inequality can be rewritten as a higher-dimensional linear matrix inequality in $\Sigma^Z$ and $G$ as follows:
\begin{eqnarray}\label{SigmaV_pos_def}
\left[ {\begin{array}{*{20}{c}}
\Sigma^Z    &    G\\
G^\top      &   (\Sigma^Y)^{-1}
\end{array}} \right] > 0.
\end{eqnarray}
When we solve the complete optimization problem, we apply inequality \eqref{SigmaV_pos_def} to enforce that the optimal $\Sigma^Z$ and $G$ lead to a positive definite $\Sigma^V$.\\
Finally, given $(\Sigma^Z,\Sigma^S,\Sigma^{S,Z})$ from Lemma \ref{stackedDist}, we can write the cost $I[S;Z]$ in terms of covariance matrices as follows:
\begin{subequations}\label{mutualinfcomplete}
\begin{align}
&I[S;Z] = h[ S] + h[ Z] - h[S,Z]\label{eq3a}\\[1mm]
& = \frac{1}{2} \log \frac{{\det \left( {\Sigma^S} \right)\det \left( {\Sigma^Z} \right)}}{{\det \left( {\Sigma^{Z,S}} \right)}}\label{eq3d}\\
& = \frac{1}{2} \log{\det \left( {\Sigma^S} \right)} - \frac{1}{2} \log{\det  {\left( {\Sigma}^S - {{\Sigma}^{ZS}}^\top {{\Sigma}^ Z}^{-1} {\Sigma}^{ZS} \right),}}\label{eq3f}
\end{align}
\end{subequations}
where \eqref{eq3d}-\eqref{eq3f} follow from standard determinant and logarithm formulas. We prove in the following lemma that minimizing the cost function $I[S;Z]$ using $(G,{{\Sigma }^Z})$ as optimization variables is equivalent to solving a convex program subject to some Linear Matrix Inequalities (LMI) constraints.

\begin{lemma}\label{mutualinformationcov}
Minimizing $I[S;Z]$ in \eqref{mutualinfcomplete} is equivalent to solving the following convex program:
\begin{eqnarray}
\left\{\begin{aligned}
	&\min_{{\Pi},\Sigma^{Z},G}\
      - \log{\det \left({\Pi} \right)} \label{finalcost_program},\\[1mm]
    &\hspace{4mm}\text{\emph{s.t. }} \Pi \geq \mathbf{0}, \begin{bmatrix}
\Sigma^S - \Pi  & (G {\Sigma}^{YS})^\top\\
G {\Sigma}^{YS} & \Sigma^Z
\end{bmatrix} \geq 0.
\end{aligned}\right.
\end{eqnarray}
\end{lemma}
\textbf{\emph{Proof}}:
Consider the expression for $I[S;Z]$ in \eqref{eq3f}. Due to monotonicity of the determinant function and the fact that $\Sigma^S$ is independent of the design variables, minimizing \eqref{eq3f} is equivalent to
\begin{eqnarray}
\left\{\begin{aligned}
	&\min_{{\Pi},\Sigma^{Z},G}\
      - \log{\det \left({\Pi} \right)} \label{epigraphcost}\\[1mm]
    &\hspace{4mm}\text{s.t. } \mathbf{0} < {\Pi} \le  {\left( {\Sigma}^S - {{\Sigma}^{ZS}}^\top {{\Sigma}^ Z}^{-1} {\Sigma}^{ZS} \right).} \label{inequalityofcost}
\end{aligned}\right.
\end{eqnarray}
The inequality term in \eqref{inequalityofcost} can be rewritten using Schur complements properties \cite{zhang2006schur} and \eqref{CovZS} as
\begin{eqnarray}
&\left[ {\begin{array}{*{20}{c}}
{\Sigma^S} - {\Pi} & (G {\Sigma}^{YS})^\top\\
{G{\Sigma}^{YS}}&{{\Sigma}^Z}
\end{array}} \right] \ge 0 \, {,{\Pi} > 0}. \label{finalcost}
\end{eqnarray}
Combining \eqref{inequalityofcost} and \eqref{finalcost}, we can conclude that minimizing $I[S;Z]$ is equivalent to solving the convex program in \eqref{finalcost_program}.
\hfill $\blacksquare$\\
By Lemma \ref{mutualinformationcov}, minimizing the cost in \eqref{problem1} is equivalent to solving the convex program in \eqref{finalcost_program}. Therefore, if the distortion constraint, $E[||W(Z - Y)||^2]<\epsilon$, is convex in the decision variables, we can employ off-the-shelf optimization algorithms to find optimal distorting mechanisms being the solution to Problem $1$.
\subsection{Distortion Constraint: Formulation and Convexity}

\begin{lemma}\label{constraint}
$E[||W(Z - Y)||^2]$ is a convex function of ${{\Sigma }^Z}$ and $G$, and can be written as follows:
\begin{align}\label{distortion}
E[||W(Z - Y)||^2] &= \text{\emph{tr}}[W^\top(\Sigma^Z + \Sigma^Y - 2\Sigma^Y G)W]\notag\\[1mm]
&+ {{\mu_Y}^\top} (G - I)^\top W^\top W  (G - I)  {\mu_Y}.
\end{align}
\end{lemma}
\textbf{\emph{Proof}}:
The expectation of the quadratic form, $E[||W(Z - Y)||^2]$, can be written (see \cite{seber2012linear} for details) in terms of the mean and covariance of the error $\Delta := W(Z - Y)$:
\begin{align}
&E[||W(Z - Y)||^2] = \text{tr}[\Sigma^\Delta] + (\mu^{\Delta})^\top \mu^\Delta \label{eq4b}.
\end{align}
with covariance $\Sigma^\Delta := E[(\Delta - \mu^{\Delta})(\Delta - \mu ^{\Delta})^\top]$ and mean $\mu ^{\Delta} := E[\Delta]$. Given the distortion mechanism \eqref{eq2}, we can write $\Delta = W((G - I) Y + V)$, and because $Y$ and $V$ are independent by assumption, $\Sigma^\Delta = W^\top (G - I)^\top \Sigma^Y (G - I) W + W^\top {{\Sigma }^V W} $ and $\mu ^\Delta = W (G - I) \mu_Y$. Then, having $\Sigma^V$ in terms of $\Sigma^Z$ in \eqref{SigmaZ}, $\Sigma^\Delta$ can be written in terms of $\Sigma^Z$ and $G$ as $\Sigma^\Delta =  W^\top(\Sigma^Z + \Sigma^Y -G^\top \Sigma^Y - \Sigma^Y G)W$. Up to this point, we have written both $\mu^{\Delta}$ and $\Sigma^\Delta$ in terms of the design variables, ${{\Sigma }^Z}$ and $G$. Using these expressions and \eqref{eq4b}, we can conclude \eqref{distortion}. \hfill $\blacksquare$
\begin{remark}
The distortion constraint $E[||W(Z - Y)||^2] \le \epsilon$ is nonlinear in the design variables, because the distortion metric \eqref{distortion} is linear in ${{\Sigma }^Z}$ and quadratic in $G$. However, we can construct an equivalent linear constraint using standard Schur complement properties. It can be proved (see \emph{\cite{zhang2006schur}} for details) that $E[||W(Z - Y)||^2] \le \epsilon$ in \eqref{distortion} is equivalent to the following LMI in $\Sigma^Z$ and $G$:
\begin{equation}
\begin{aligned}
\left\{
\begin{array}{ll}
&\begin{bmatrix} \theta_Y & {{\mu_Y}^\top}(G - I)^\top W^\top  \\ W(G - I) {\mu_Y} & I \end{bmatrix} \ge 0,\\[6mm]
&\theta_Y := \epsilon - \text{\emph{tr}}[ W^\top (\Sigma^Z + \Sigma^Y - 2\Sigma^Y G)W ].
\end{array}
\right.
\end{aligned}
\end{equation}

\end{remark}
By Lemma \ref{mutualinformationcov}, Lemma \ref{constraint}, and Remark $1$, the cost function $I[S;Z]$ and distortion constraint $E[||W(Z - Y)||^2] \leq \epsilon$ can be written in terms of convex functions in the optimization variables ${{\Sigma }^Z}$ and $G$. In what follows, we pose the complete nonlinear convex program to solve Problem $1$.
\begin{table}
\noindent\rule{\hsize}{1pt}
\begin{equation} \label{eq:convex_optimization15}
\begin{aligned}
	&\min_{{\Pi}, \Sigma^{Z}, G}  - \log\det [\Pi],\\[1mm]
    &\text{ s.t. }\left\{\begin{aligned}
    &{\Pi} \ge 0,\\ &\left[ {\begin{array}{*{20}{c}}
\Sigma^S - \Pi  & (G\Sigma^{YS})^\top\\
G\Sigma^{YS} & \Sigma^Z
\end{array}} \right] \ge 0, \\[1mm]
&\left[ {\begin{array}{*{20}{c}}
\Sigma^Z    &    G\\
G^\top      &   (\Sigma^Y)^{-1}
\end{array}} \right] > 0, \\[1mm]
&\begin{bmatrix} \theta_Y & {{\mu^Y}^\top}(G - I)^\top W^\top \\ W(G - I) {\mu^Y} & I \end{bmatrix} \ge 0,\\[1mm]
&\theta_Y := \epsilon - \text{\emph{tr}}[ W^\top (\Sigma^Z + \Sigma^Y - 2\Sigma^Y G)W ]. \end{aligned}\right.
\end{aligned}
\end{equation}
\noindent\rule{\hsize}{1pt}\end{table}
\begin{theorem}\label{th3}
Consider the distorting mechanism \eqref{eq2}, desired output distortion level ${\epsilon} \in {\mathbb{R}^+}$, output distortion weight ${W} \in {\mathbb{R}^{ {n_y} \times  {n_y}}}$, covariance $\Sigma^{S}$, cross-covariance $\Sigma^{YS}$, and the mean and covariance of $Y$, $\mu^{Y}$ and $\Sigma^{Y}$, respectively. Then, the optimization variables ${{\Sigma }^Z}$ and $G$ that minimize $I[S;Z]$ subject to distortion constraint $E[||W(Z - Y)||^2] \leq \epsilon$, as in Problem 1, can be found by solving the convex program in \eqref{eq:convex_optimization15}.
\end{theorem}
\emph{\textbf{Proof:}} The expressions for the cost and constraint and convexity (linearity) of them follow from Lemma \ref{mutualinformationcov}, Lemma \ref{constraint}, and Remark $1$.  \hfill $\blacksquare$\\

\begin{remark}
Theorem 1 provides a tool to find optimal privacy-preserving mechanisms when the prior distributions are jointly Gaussian. In the following proposition, we prove that we can allow for priors that are not necessarily Gaussian, as long as they are log-concave \emph{\cite{prekopa1980logarithmic}}. The latter generalizes significantly the class of databases that Theorem 1 can deal with. In particular, we prove that if $S$ and $Z$ follow non-Gaussian log-concave distributions, their mutual information is upper-bounded by an affine function of the mutual information between $S_{\mathcal{N}}$ and $Z_{\mathcal{N}}$, where $S_{\mathcal{N}}$ and $Z_{\mathcal{N}}$ denote Gaussian random vectors with the same mean and covariance matrix as $S$ and $Z$, respectively. Examples of log-concave distributions are normal, exponential, uniform, Laplace, chi, beta, and logistic distributions. We refer the reader to \emph{\cite{LogConcave}} for a comprehensive list and properties of log-concave distributions.
\end{remark}

\begin{proposition}\label{mutualinformationupper}
Let $S$ and $Z$ be jointly distributed following log-concave multivariate joint probability distribution. Their mutual information $I[S;Z]$ can be upper-bounded as:
\begin{eqnarray}
   I[S;Z] \le I[S_{\mathcal{N}};Z_{\mathcal{N}}] +  C_n,\label{eqmutualinfupper2}
\end{eqnarray}
where $C_n = \frac{1}{2}\log(2\pi e c(n))^n$, $n=n_s + n_y$, $c(n)=\frac{e^{2} n^{2}}{4 \sqrt{2}(n+2)}$, and $S_{\mathcal{N}}$ and $Z_{\mathcal{N}}$ denote Gaussian random vectors with the same covariance matrices as $S$ and $Z$.
\end{proposition}
\textbf{\emph{Proof}}:
First, we need to prove that for $S$ and $Y$ following log-concave distribution, $Z$ also follows a log-concave distribution. From the fact that the log-concavity of distribution is preserved under affine transformation, we can conclude that $GY$ is log-concave (see \cite{dharmadhikari1988unimodality}, Lemma 2.1). Then, based on Hoggar’s theorem, the sum of two independent log-concave random variables is itself log-concave, which resulted in log-concavity of $Z$ as a summation of $GY$ and $V$ \cite{hoggar1974chromatic,johnson2006preservation}.\\
Mutual information $I\left[S;Z \right]$ can be written in terms of differential entropies as $I[S;Z] = h[S] + h[Z] - h[S,Z]$. Moreover, because Gaussian distributions maximize entropy for a given covariance matrix \cite{Cover}, we can write:
\begin{equation}
    \left\{\begin{aligned}
h[S] &\le h[S_{\mathcal{N}}],\\
h[Z] &\le h[Z_{\mathcal{N}}],
\end{aligned}\right. \label{inequalityentropy}
\end{equation}
where $S_{\mathcal{N}}$ and $Z_{\mathcal{N}}$ denote Gaussian random vectors with the same covariance matrices as $S$ and $Z$.
The authors in \cite{marsiglietti2018lower} prove that the entropy of random vectors following log-concave distributions \cite{wellner2012log} can be lower bounded as:
\begin{equation}
    h[S,Z] \ge \frac{n}{2} \log \frac{\left|\Sigma^{S,Z}\right|^{1 / n}}{c(n)},\label{entropylower}
\end{equation}
where $n=n_s + n_z$ and $c(n)=\frac{e^{2} n^{2}}{4 \sqrt{2}(n+2)}$. Expanding the lower bound in \eqref{entropylower}, we obtain
\begin{align}\label{lowerboundformula}
   \frac{n}{2} \log \frac{\left|\Sigma^{S,Z}\right|^{1 / n}}{c(n)} &=\frac{n}{2}\left(\frac{1}{n} \log \operatorname{det} \Sigma^{S,Z}-\log c(n)\right)\nonumber \\
   &=\frac{1}{2} \log \operatorname{det} \Sigma^{S,Z} - \frac{n}{2} \log c(n).
\end{align}
Let $(S_{\mathcal{N}},Z_{\mathcal{N}})$ be jointly Gaussian with covariance $\Sigma^{S,Z}$. Then, we can write its differential entropy as $h[S_{\mathcal{N}},Z_{\mathcal{N}}] = \frac{1}{2}\log \det \left( \Sigma^{S,Z} \right) + \frac{n}{2} + \frac{n}{2}\log(2\pi)$ (see Theorem $8.4.1$ in \cite{Cover}). Therefore, using the latter formula for $h[S_{\mathcal{N}},Z_{\mathcal{N}}]$ and \eqref{lowerboundformula}, we can write the following 
\begin{equation}
     \frac{n}{2} \log \frac{\left|\Sigma^{S,Z}\right|^{1 / n}}{c(n)}= h[S_{\mathcal{N}},Z_{\mathcal{N}}] - \frac{n}{2} - \frac{n}{2}\log(2\pi) - \frac{n}{2} \log c(n).\label{entropylower2}
\end{equation}
Combining \eqref{inequalityentropy}, \eqref{entropylower}, and \eqref{entropylower2}, we can finally conclude that
\begin{align}
       I[S;Z] &\le h[S_{\mathcal{N}}] + h[Z_{\mathcal{N}}] - h[S_{\mathcal{N}},Z_{\mathcal{N}}]\nonumber\\ &+\frac{n}{2} + \frac{n}{2}\log(2\pi)+ \frac{n}{2} \log c(n),
\end{align}
which is equivalent to \eqref{eqmutualinfupper2}.
 \hfill $\blacksquare$

\begin{remark}\label{upperboundremark1}
Proposition 1 implies that by minimizing $I[S_{\mathcal{N}};Z_{\mathcal{N}}]$ (based on the optimization problem in Theorem 1), we can effectively decrease the mutual information $I[S;Z]$ for any $(S,Z)$ following a log-concave distribution.
\end{remark}
A tighter bound for the mutual information of log-concave random vectors can be achieved based on their maximum density. The maximum density of a random vector is calculated by its $L^\infty$ norm \cite{bobkov2011entropy}. In the following proposition, we prove that the mutual information between log-concave random vectors $S$ and $Z$ can be upper-bounded by an affine function of the mutual information of Gaussian random vectors with maximum density being the same as those of $S$ and $Z$. 

\begin{proposition}\label{mutualinformationupper2}
Let $S$ and $Z$ be jointly distributed following a log-concave multivariate joint probability distribution. For their mutual information $I[S;Z]$, one can write:
\begin{eqnarray}
  \frac{1}{n} I[S;Z] - \frac{1}{n} I[S^*_{\mathcal{N}};Z^*_{\mathcal{N}}] \le 1,\label{eqmutualinfupper}
\end{eqnarray}
where $n=n_s + n_y$, and $S^*_{\mathcal{N}}$ and $Z^*_{\mathcal{N}}$ denote multivariate normally distributed random vectors with maximum density being the same as those of $S$ and $Z$, respectively.
\end{proposition}
\textbf{\emph{Proof}}:
Mutual information $I\left[S;Z \right]$ can be written in terms of differential entropies as $I[S;Z] = h[S] + h[Z] - h[S,Z]$. Moreover, from the inequality on the entropy of log-concave distributed random vectors, we can conclude that (see Theorem $I.1$ in \cite{bobkov2011entropy}):
\begin{equation}
    \left\{\begin{aligned}
\frac{1}{n_s} h[S] &\le \frac{1}{n_s}  h[S^*_{\mathcal{N}}] + \frac{1}{2},\\
\frac{1}{n_y} h[Z] &\le \frac{1}{n_y} h[Z^*_{\mathcal{N}}] + \frac{1}{2},\\
\frac{1}{n} h[S,Z] &\ge \frac{1}{n} h[S^*_{\mathcal{N}},Z^*_{\mathcal{N}}] - \frac{1}{2},
\end{aligned}\right. \label{inequalityentropy2}
\end{equation}
where $n=n_s + n_y$, and $S^*_{\mathcal{N}}$ and $Z^*_{\mathcal{N}}$ denote multivariate normally distributed random vectors with maximum density being the same as those of $S$ and $Z$, respectively.\\
Combining all these inequalities in \eqref{inequalityentropy2}, we can conclude that
\begin{align}
       \frac{1}{n} I[S;Z] &\le \frac{1}{n}(h[S^*_{\mathcal{N}}] + h[Z^*_{\mathcal{N}}] - h[S^*_{\mathcal{N}},Z^*_{\mathcal{N}}]) + 1,
\end{align}
which is equivalent to \eqref{eqmutualinfupper}.
 \hfill $\blacksquare$

\begin{remark}\label{upperboundremark}
In Proposition 2 we prove that the difference between $I[S;Z]$ and $I[S^*_{\mathcal{N}};Z^*_{\mathcal{N}}]$ can be upper-bounded by $n$. This upper bound is still conservative since it covers all log-concave distributed random vectors. However, the authors in \cite{bobkov2011entropy} observe that every log-concave random vector has approximately the same entropy per coordinate as a related Gaussian vector (which resulted in approximately the same mutual information) and log-concave distributions resemble log-concave distributions. Therefore, we can imply that by minimizing $I[S^*_{\mathcal{N}};Z^*_{\mathcal{N}}]$ (based on the optimization problem in Theorem 1), we can effectively decrease the mutual information $I[S;Z]$ for any $(S,Z)$ following log-concave distribution.
\end{remark}
\begin{figure}[!htb]
  \centering
  \includegraphics[width=3.5in]{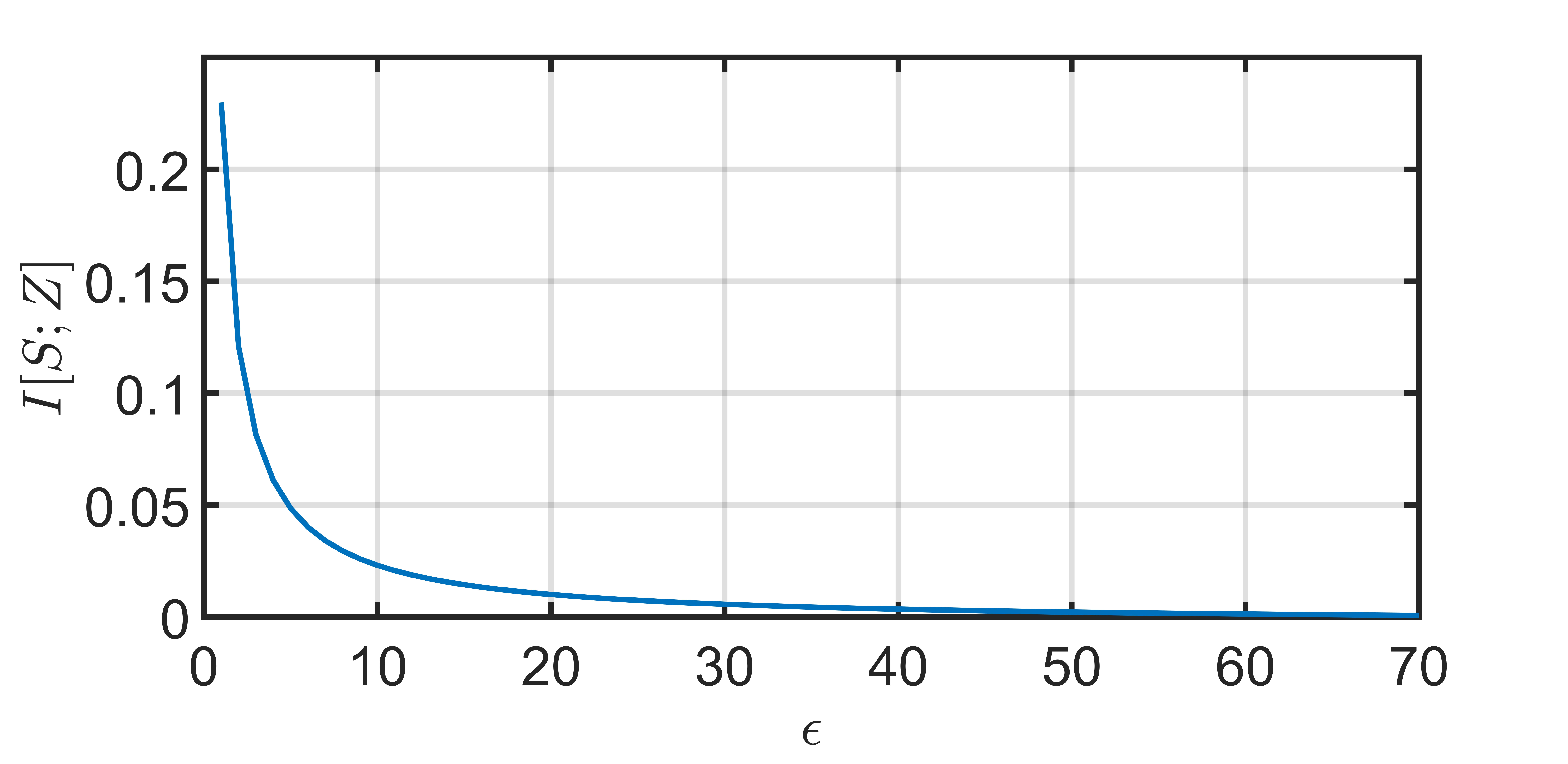}
  \caption{Cost function for increasing $\epsilon$.}\label{CostBasedEps}
\end{figure}
\section{SIMULATION RESULTS}
We consider multivariate Gaussian private $S$ and disclosed $Y$ with $n_s = n_y = 3$, and randomly selected means and covariances $\mu_S$, $\mu_Y$, $\Sigma^S>0$, and $\Sigma^Y>0$. We first show the effect of the distortion level $\epsilon$ on the amount of information leakage, calculated by the optimal cost function $I[S;Z]$, in Figure \ref{CostBasedEps}. As expected, the value of the cost decreases monotonically as we increase the maximum allowed distortion $\epsilon$. This figure illustrates that by increasing the distortion level, the decreasing rate of the optimal cost is reduced (we actually have exponential decay). So the best privacy-distortion trade-off is reached for small $\epsilon$.
\begin{figure}[ht]
\centering
\begin{subfigure}{.5\textwidth}
  \centering
  \includegraphics[width=3.5in]{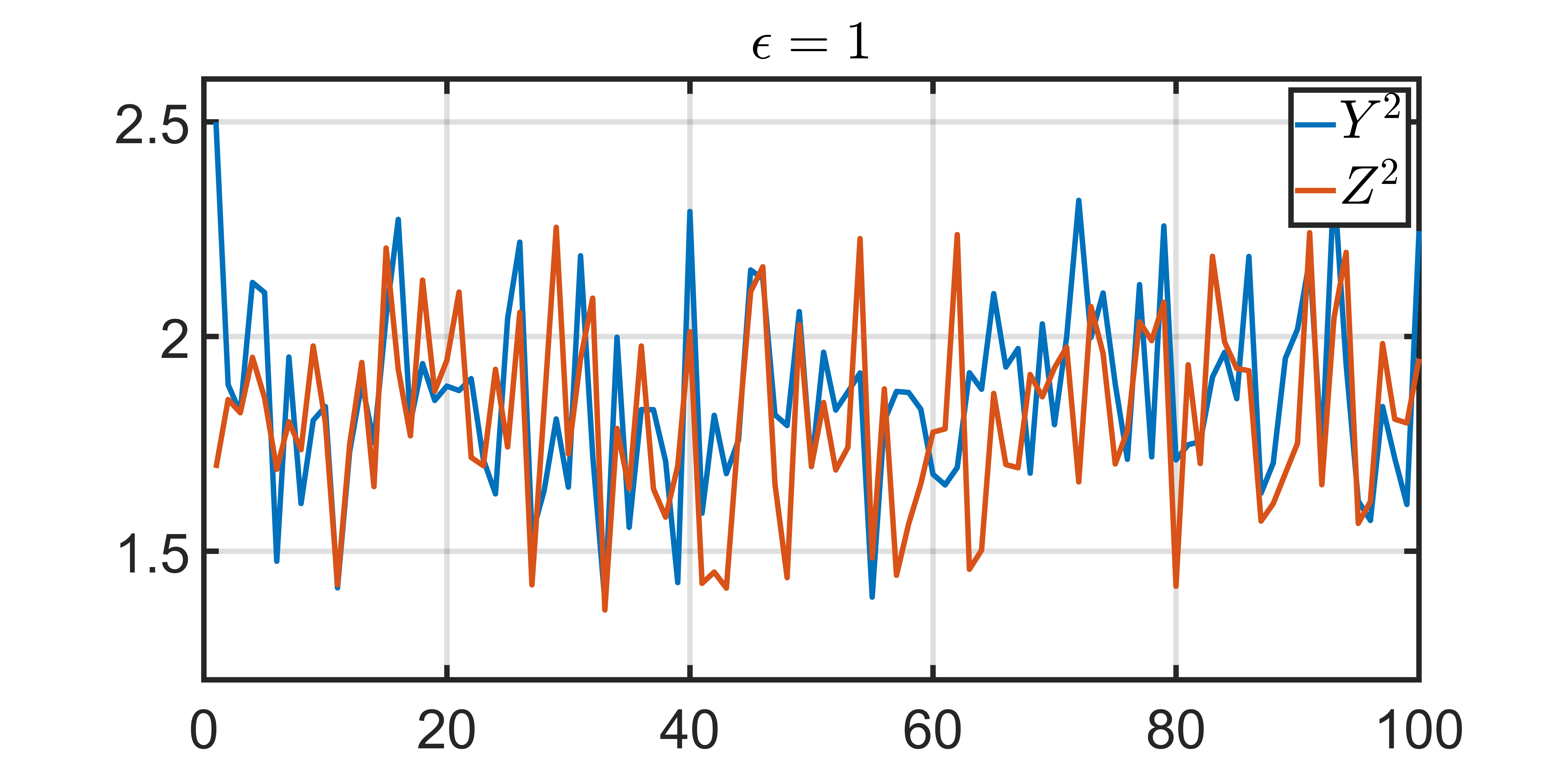}
  \label{fig:sub-first}
\end{subfigure}
\begin{subfigure}{.5\textwidth}
  \centering
  \includegraphics[width=3.5in]{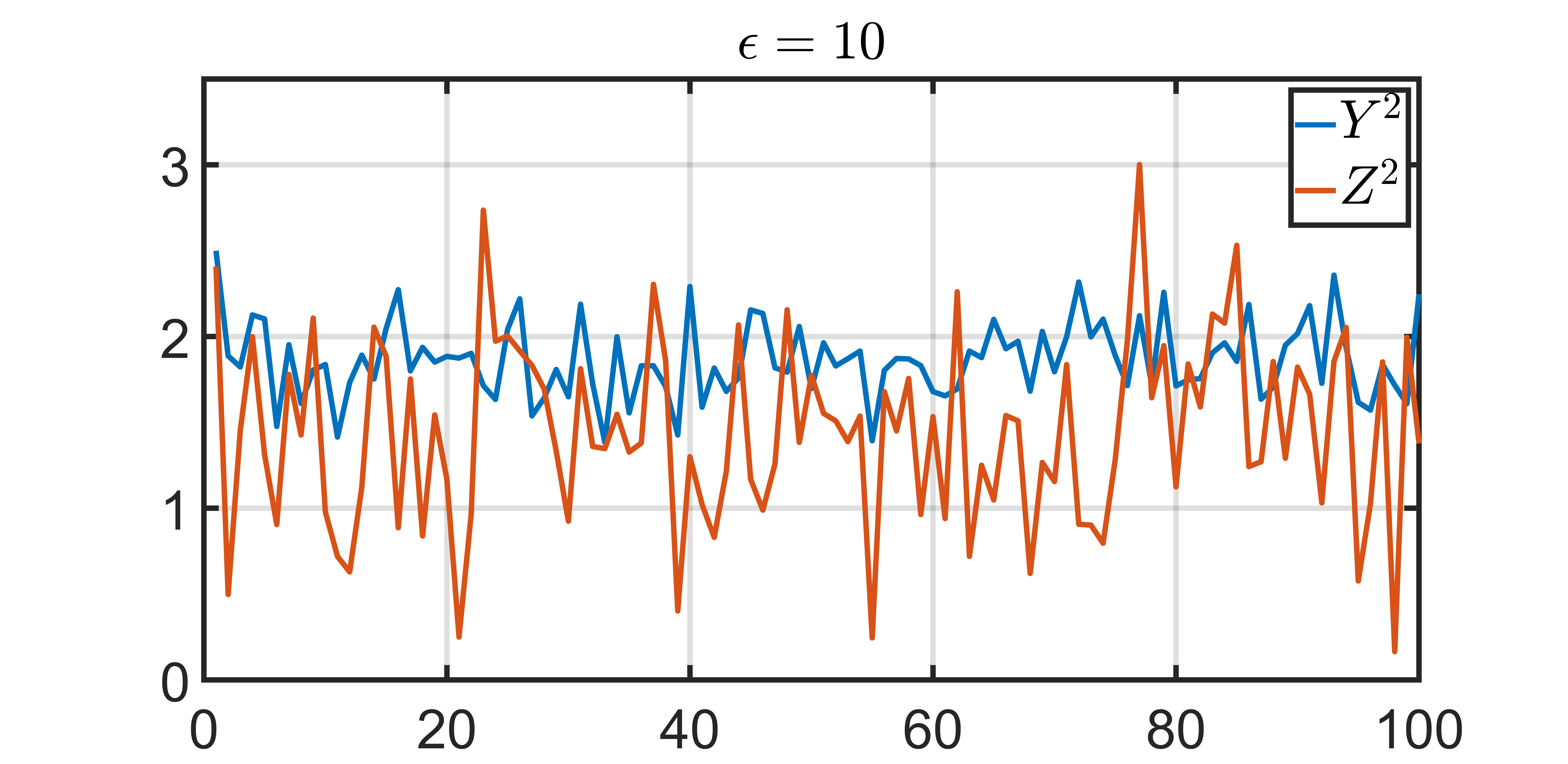}
  \label{fig:sub-second}
\end{subfigure}
\begin{subfigure}{.5\textwidth}
  \centering
  \includegraphics[width=3.5in]{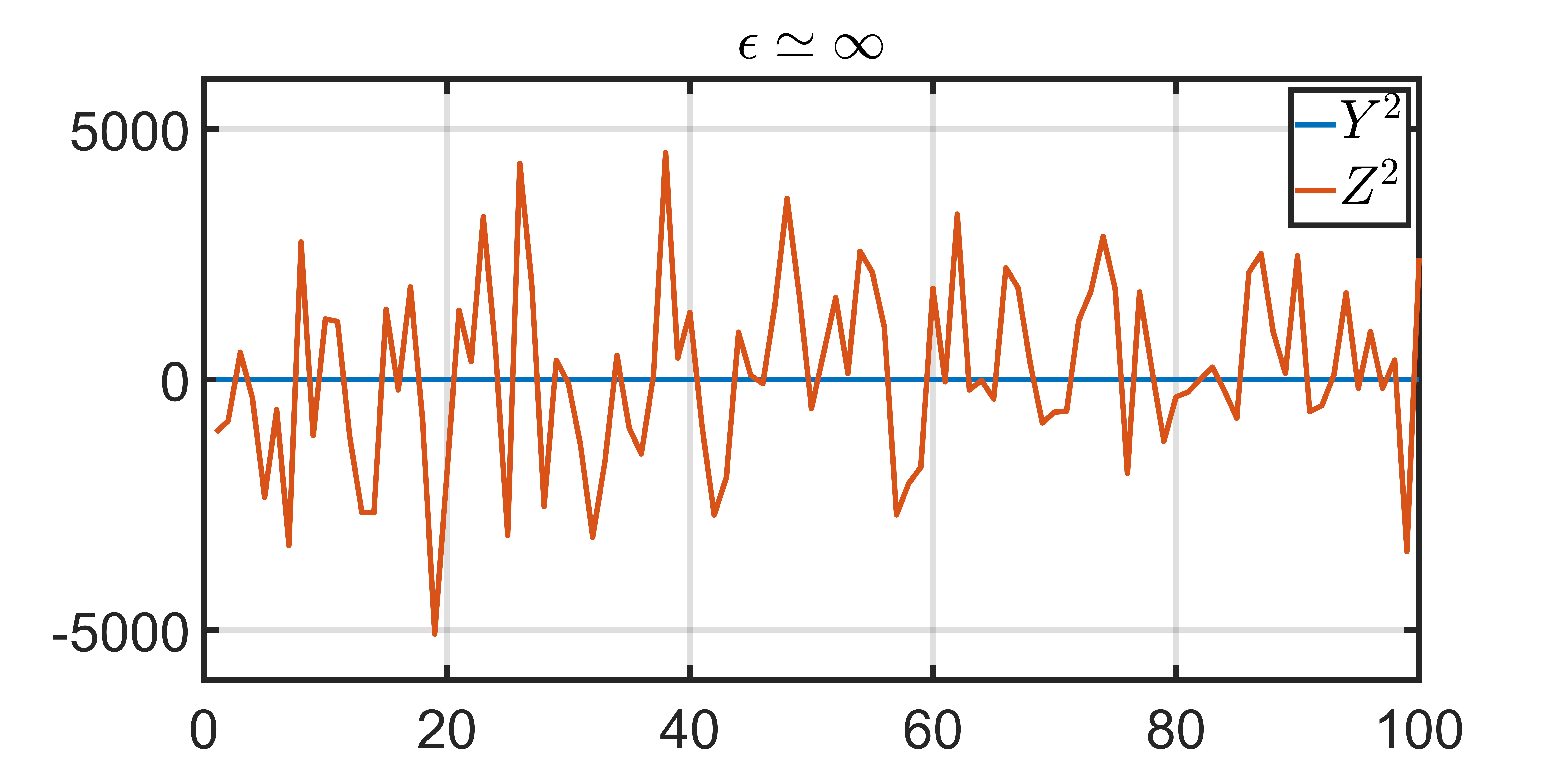}
  \label{fig:sub-second}
\end{subfigure}
\caption{Comparison between the second element of $Y$ and $Z$ for different distortion levels, $\epsilon =1, 10, \infty$.}
\label{YZ}
\end{figure}
\begin{figure}[ht]
\centering
\begin{subfigure}{.5\textwidth}
  \centering
  \includegraphics[width=3.5in]{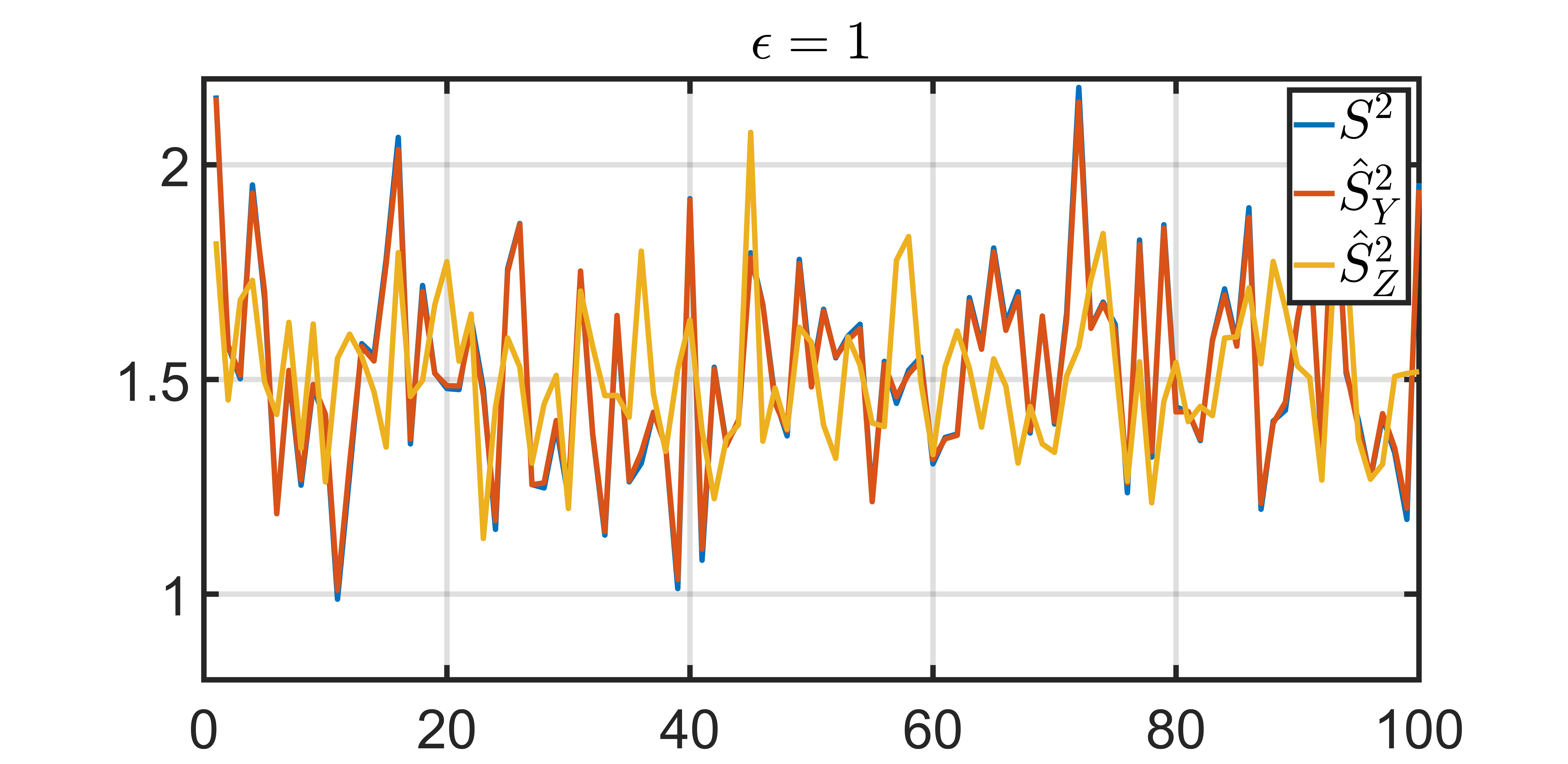}
  \label{S-sub-first}
\end{subfigure}
\begin{subfigure}{.5\textwidth}
  \centering
  \includegraphics[width=3.5in]{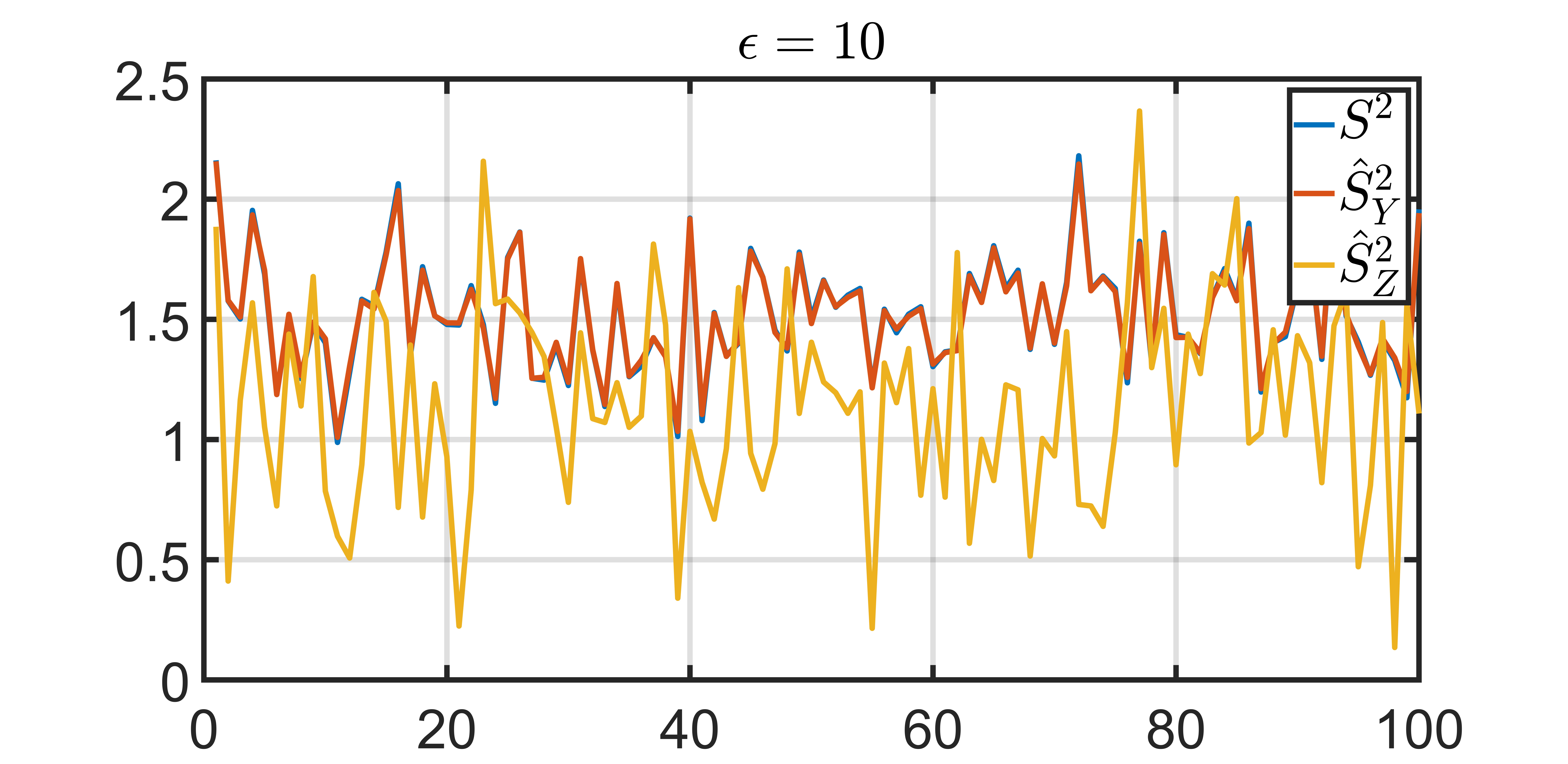}
  \label{S-sub-second}
\end{subfigure}
\begin{subfigure}{.5\textwidth}
  \centering
  \includegraphics[width=3.5in]{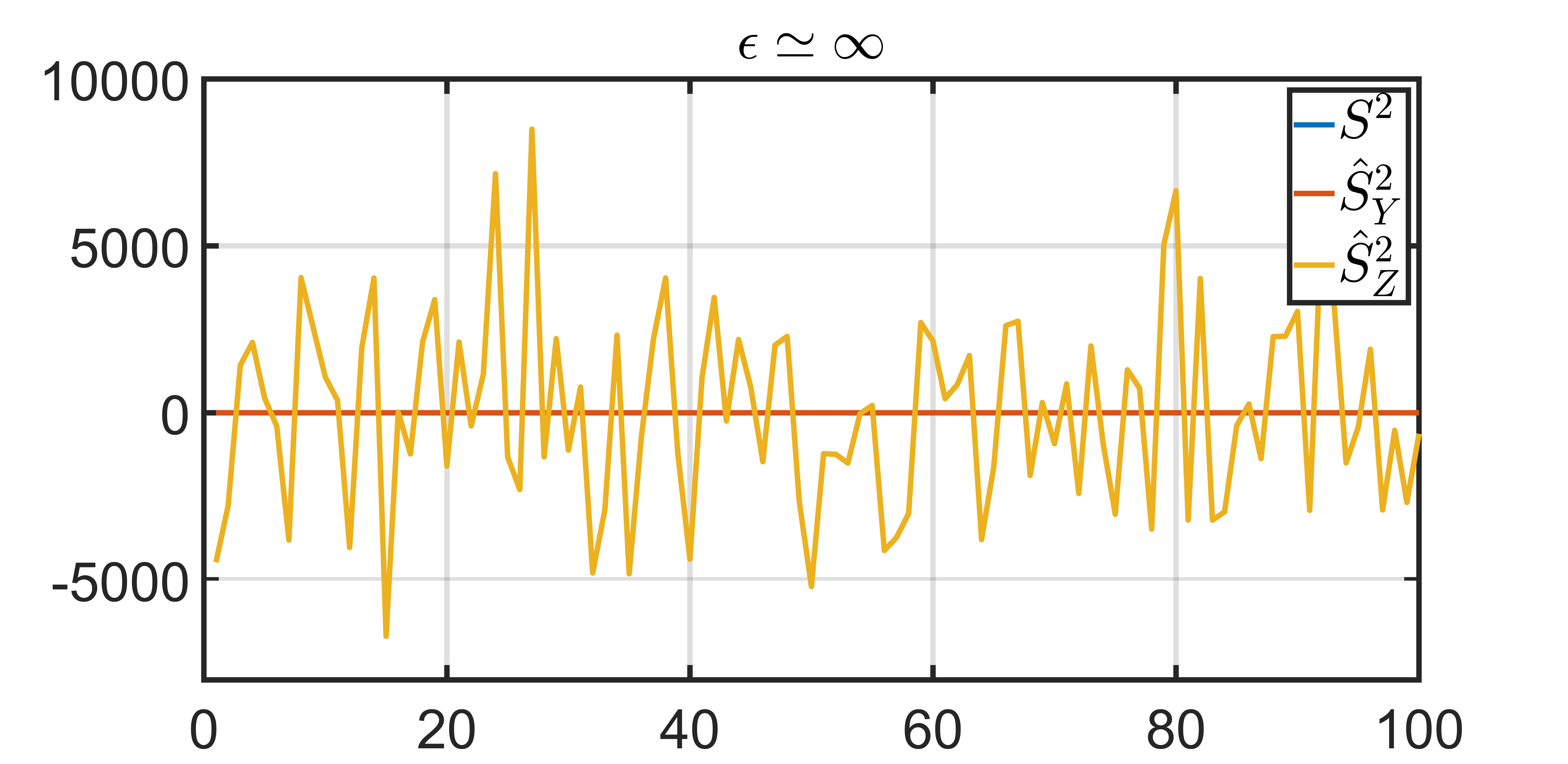}
  \label{S-sub-3rd}
\end{subfigure}
\caption{Comparison between the second element of the private data, $S^2$, its estimation without distorting $Y$, ${\hat S^2_{Y}}$, and its estimation given distorted data $Z$, ${\hat S^2_{Z}}$, for different distortion levels $\epsilon =1, 10, \infty$.}
\label{estimatedS}
\end{figure}
\begin{figure}[!htb]
  \centering
  \includegraphics[width=3.5in]{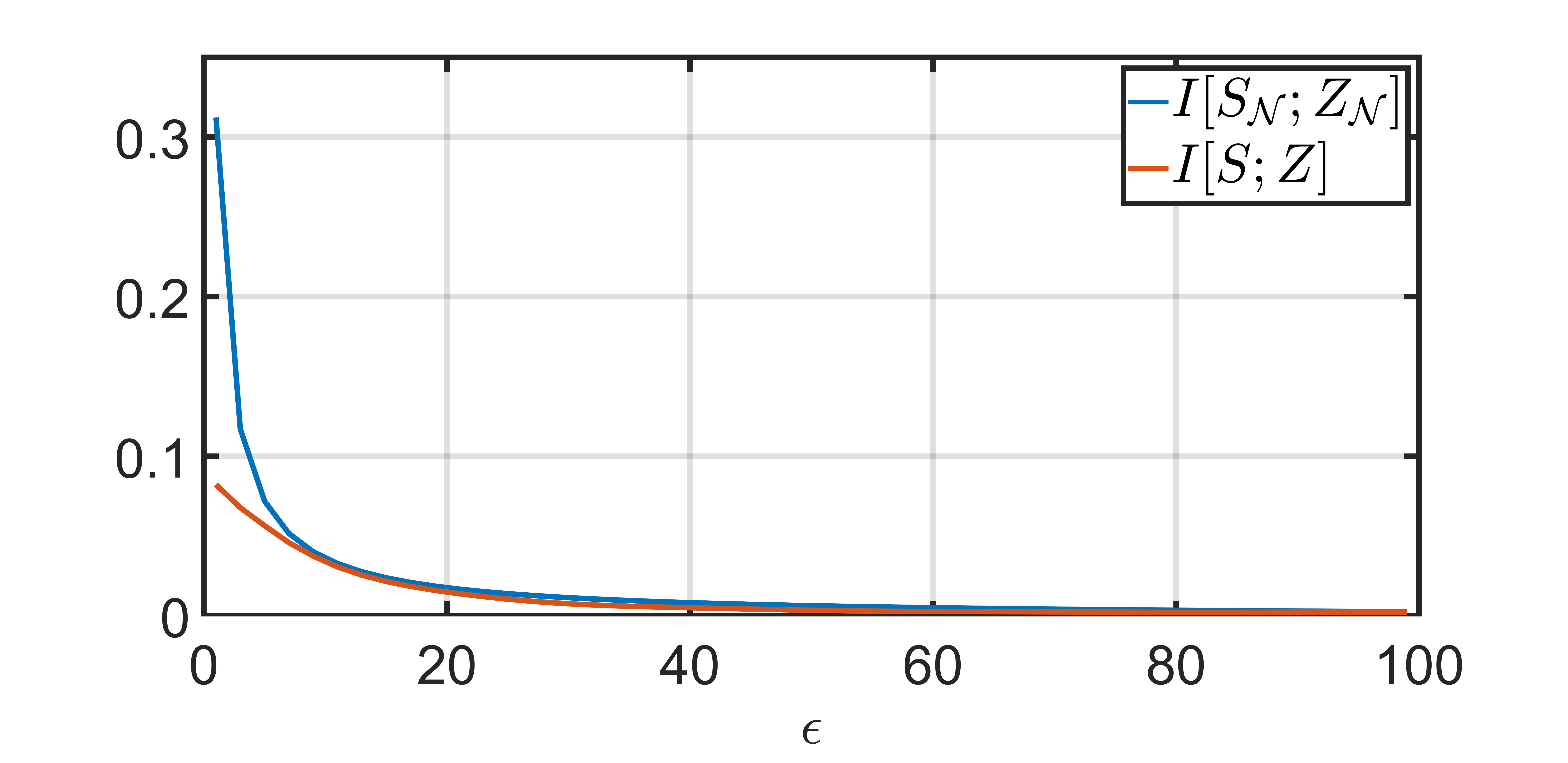}
  \caption{Comparison between the mutual information of $S$ and $Z$, with Laplacian $S$ and $Y$, $I[S;Z]$, and its upper bound, which is based on the mutual information of Gaussian vectors $S_{\mathcal{N}}$ and $Z_{\mathcal{N}}$, $I[S_{\mathcal{N}};Z_{\mathcal{N}}]$, for increasing $\epsilon$.}
\label{laplacedistribution}
\end{figure}\\
The effect of the optimal distorting mechanisms is illustrated in Figure \ref{YZ} for different levels of distortion $\epsilon = \{ 1,10,\infty \}$. Here, we contrast realizations of the second elements of $Y$ and distorted $Z$. The value of $\epsilon = \infty$ means that the optimization problem in \eqref{eq:convex_optimization15} is solved without considering the distortion constraint between $Z$ and $Y$. The difference between $Y$ and $Z$ increases by increasing the distortion level.\\
Next, in Figure \ref{estimatedS}, we show realizations of the second element of the private $S$ and its MMSE estimate, ${{\hat S}_{Y}}^2$, constructed from realizations of the undistorted $Y$ and its MMSE estimate given realizations of the distorted $Z$, ${{\hat S^2}_{Z}}$. We consider different levels of distortion, $\epsilon = \{ 1,10,\infty \}$. In Figure \ref{estimatedS}, we observe that by randomizing $Y$, we can prevent accurate estimation of $S$. Because by increasing the distortion level $\epsilon$, the accuracy of the estimation decreases.\\
Finally, in Figure \ref{laplacedistribution}, we show that, as discussed in Remark \ref{upperboundremark1} and Remark \ref{upperboundremark}, for non-Gaussian random vectors $S$ and $Z$ following joint log-concave distributions, their mutual information can be upper-bounded by affine functions of the mutual information of some Gaussian random vectors. The upper bounds are based on the mutual information of Gaussian random vectors $S_{\mathcal{N}}$ and $Z_{\mathcal{N}}$ with the same mean and covariance matrices as $S$ and $Z$, and also the mutual information of $S^*_{\mathcal{N}}$ and $Z^*_{\mathcal{N}}$ denote multivariate normally distributed random vectors with maximum density being the same as those of $S$ and $Z$, respectively. We use $S$ and $Y$ with joint Laplace distribution and solve the optimization problem in \eqref{eq:convex_optimization15} for $I[S_{\mathcal{N}};Z_{\mathcal{N}}]$ to find the optimal distorting parameters $G$ and $\Sigma^V$, for increasing amount of distortion level $\epsilon$. 
Since the maximum density of Gaussian random vectors and Laplace random vectors are the same for the same mean and covariances, we can conclude that for Laplace variables $S^*_{\mathcal{N}}=S_{\mathcal{N}}$, $Z^*_{\mathcal{N}}=Z_{\mathcal{N}}$, and therefore $I[S^*_{\mathcal{N}};Z^*_{\mathcal{N}}]=I[S_{\mathcal{N}};Z_{\mathcal{N}}]$.
We calculate $I[S_{\mathcal{N}};Z_{\mathcal{N}}]+C_n$ and $I[S^*_{\mathcal{N}};Z^*_{\mathcal{N}}]+n$ as the upper bounds of $I[S;Z]$, where in this case study, $C_n=18.6204$ and $n=6$. Then, we calculate the mutual information between $S$ and $Z$ \cite{MImatlab}, where $S$ and $Y$ follow Laplace distributions with the same mean and covariance as $S_{\mathcal{N}}$ and $Y_{\mathcal{N}}$ and $Z=GY +V$ (with the same distortion variables as Gaussian case). As can be seen in Figure \ref{laplacedistribution}, $I[S;Z]$ is even upper-bounded by $I[S_{\mathcal{N}};Z_{\mathcal{N}}]$ without the constant terms $n$ and $C_n$. It implies that by solving the optimization problem for Gaussian data, the information leakage for Laplace data, which is given by $I[S;Z]$, is decreasing for increasing $\epsilon$, and is upper-bounded by the mutual information $I[S^*_{\mathcal{N}};Z^*_{\mathcal{N}}]$ and $I[S_{\mathcal{N}};Z_{\mathcal{N}}]$. Furthermore, it is apparent that even if the Gaussian upper bounds is loose for these distributions of $S$ and $Z$, the information leakage $I[S;Z]$ is still decreasing for increasing $\epsilon$. This implies that minimizing $I[S_{\mathcal{N}};Z_{\mathcal{N}}]$ effectively decreases the mutual information $I[S;Z]$ for any log-concave $S$ and $Z$.
\section{CONCLUSIONS}
This paper proposes a comprehensive mathematical framework for synthesizing distorting mechanisms to minimize the information leakage caused by public/unsecured communication networks for a class of stochastic databases. We have introduced a class of linear Gaussian distorting mechanisms to randomize data before transmission to prevent adversaries from precisely estimating the private part of the database.\\
Furthermore, for the class of data under study, we have thoroughly characterized an information-theoretic metric (mutual information) to quantify the information between private data and disclosed data for a class of worst-case eavesdropping adversaries.\\
Finally, given the maximum level of distortion tolerated by a particular application, we have provided tools (in terms of convex programs) for designing optimal (in terms of maximizing privacy) distorting mechanisms. We have presented simulation results to illustrate the performance of our tools.
\bibliographystyle{IEEEtran}
\bibliography{conference_101719.bbl}

\begin{thebibliography}{10}
\providecommand{\url}[1]{#1}
\csname url@samestyle\endcsname
\providecommand{\newblock}{\relax}
\providecommand{\bibinfo}[2]{#2}
\providecommand{\BIBentrySTDinterwordspacing}{\spaceskip=0pt\relax}
\providecommand{\BIBentryALTinterwordstretchfactor}{4}
\providecommand{\BIBentryALTinterwordspacing}{\spaceskip=\fontdimen2\font plus
\BIBentryALTinterwordstretchfactor\fontdimen3\font minus
  \fontdimen4\font\relax}
\providecommand{\BIBforeignlanguage}[2]{{%
\expandafter\ifx\csname l@#1\endcsname\relax
\typeout{** WARNING: IEEEtran.bst: No hyphenation pattern has been}%
\typeout{** loaded for the language `#1'. Using the pattern for}%
\typeout{** the default language instead.}%
\else
\language=\csname l@#1\endcsname
\fi
#2}}
\providecommand{\BIBdecl}{\relax}
\BIBdecl

\bibitem{Poor1}
S.~R. {Rajagopalan}, L.~{Sankar}, S.~{Mohajer}, and H.~V. {Poor}, ``Smart meter
  privacy: A utility-privacy framework,'' in \emph{Proceedings of the IEEE
  International Conference on Smart Grid Communications (SmartGridComm)}, 2011,
  pp. 190--195.

\bibitem{Poor2}
O.~{Tan}, D.~{G\"{u}nd\"{u}z}, and H.~V. {Poor}, ``Increasing smart meter
  privacy through energy harvesting and storage devices,'' \emph{IEEE Journal
  on Selected Areas in Communications}, vol.~31, pp. 1331--1341, 2013.

\bibitem{Huang:2014:CDP:2566468.2566474}
Z.~Huang, Y.~Wang, S.~Mitra, and G.~E. Dullerud, ``On the cost of differential
  privacy in distributed control systems,'' in \emph{Proceedings of the 3rd
  International Conference on High Confidence Networked Systems}, 2014, pp.
  105--114.

\bibitem{Farokhi1}
F.~Farokhi and H.~Sandberg, ``Optimal privacy-preserving policy using
  constrained additive noise to minimize the {f}isher information,'' in
  \emph{Proceedings of the IEEE 56th Annual Conference on Decision and Control
  (CDC)}, 2017.

\bibitem{Farokhi2}
F.~Farokhi, H.~Sandberg, I.~Shames, and M.~Cantoni, ``Quadratic {G}aussian
  privacy games,'' in \emph{Proceedings of the 54th IEEE Conference on Decision
  and Control (CDC)}, 2015, pp. 4505--4510.

\bibitem{Pappas}
F.~Miao, Q.~Zhu, M.~Pajic, and G.~J. Pappas, ``Coding sensor outputs for
  injection attacks detection,'' in \emph{Proceedings of the 53rd Annual
  Conference on Decision and Control (CDC)}, 2014, pp. 5776--5781.

\bibitem{Jerome1}
J.~L. Ny and G.~J. Pappas, ``Differentially private filtering,'' \emph{IEEE
  Transactions on Automatic Control}, vol.~59, pp. 341--354, 2014.

\bibitem{Takashi_1}
E.~Nekouei, T.~Tanaka, M.~Skoglund, and K.~H. Johansson,
  ``Information-theoretic approaches to privacy in estimation and control,''
  \emph{Annual Reviews in Control}, vol.~47, pp. 412 -- 422, 2019.

\bibitem{Takashi_3}
H.~S. Takashi~Tanaka, Mikael~Skoglund and K.~H. Johansson, ``Directed
  information as privacy measure in cloud-based control,'' technical report,
  arXiv:1705.02802 [math.OC]. \url{https://arxiv.org/abs/1705.02802}.

\bibitem{chaper_privacy_chaos}
F.~F. Murguia~C., Shames~I. and N.~D., \emph{Information-Theoretic Privacy
  Through Chaos Synchronization and Optimal Additive Noise. In: Farokhi F.
  (eds) Privacy in Dynamical Systems}.\hskip 1em plus 0.5em minus 0.4em\relax
  Singapore: Springer, 2020.

\bibitem{Carlos_Iman1}
C.~Murguia, I.~Shames, F.~Farokhi, and D.~Ne\v{s}i\'{c}, ``On privacy of
  quantized sensor measurements through additive noise,'' in \emph{Proceedings
  of the 57th IEEE Conference on Decision and Control (CDC)}, 2018.

\bibitem{Cover}
T.~M. Cover and J.~A. Thomas, \emph{Elements of Information Theory}.\hskip 1em
  plus 0.5em minus 0.4em\relax New York, NY, USA: Wiley-Interscience, 1991.

\bibitem{Dwork}
C.~Dwork, ``Differential privacy: A survey of results,'' in \emph{Theory and
  Applications of Models of Computation}.\hskip 1em plus 0.5em minus
  0.4em\relax Berlin, Heidelberg: Springer Berlin Heidelberg, 2008, pp. 1--19.

\bibitem{Dwork2}
C.~Dwork and A.~Roth, ``The algorithmic foundations of differential privacy,''
  \emph{Foundations and Trends in Theoretical Computer Science}, vol.~9, pp.
  211--407, 2014.

\bibitem{Topcu}
S.~Han, U.~Topcu, and G.~J. Pappas, ``Differentially private convex
  optimization with piecewise affine objectives,'' in \emph{Proceedings of the
  53rd IEEE Conference on Decision and Control}, 2014.

\bibitem{SORIA}
J.~Soria-Comas and J.~Domingo-Ferrer, ``Optimal data-independent noise for
  differential privacy,'' \emph{Information Sciences}, vol. 250, pp. 200 --
  214, 2013.

\bibitem{Geng}
Q.~Geng and P.~Viswanath, ``The optimal mechanism in differential privacy,'' in
  \emph{Proceedings of the IEEE International Symposium on Information Theory},
  2014, pp. 2371--2375.

\bibitem{Dullerud}
Y.~Wang, Z.~Huang, S.~Mitra, and G.~E. Dullerud, ``Entropy-minimizing mechanism
  for differential privacy of discrete-time linear feedback systems,'' in
  \emph{Proceedings of the 53rd IEEE Conference on Decision and Control}, 2014,
  pp. 2130--2135.

\bibitem{FAROKHI3}
F.~Farokhi and G.~Nair, ``Privacy-constrained communication,''
  \emph{IFAC-PapersOnLine}, vol.~49, pp. 43 -- 48, 2016.

\bibitem{Fawaz}
F.~du~Pin~Calmon and N.~Fawaz, ``Privacy against statistical inference,'' in
  \emph{Proceedings of the 50th Annual Allerton Conference on Communication,
  Control, and Computing (Allerton)}, 2012, pp. 1401--1408.

\bibitem{Fawaz2}
S.~Salamatian, A.~Zhang, F.~du~Pin~Calmon, S.~Bhamidipati, N.~Fawaz, B.~Kveton,
  P.~Oliveira, and N.~Taft, ``Managing your private and public data: Bringing
  down inference attacks against your privacy,'' \emph{IEEE Journal of Selected
  Topics in Signal Processing}, vol.~9, pp. 1240--1255, 2015.

\bibitem{Lalita}
E.~V. {Belmega}, L.~{Sankar}, and H.~V. {Poor}, ``Enabling data exchange in
  two-agent interactive systems under privacy constraints,'' \emph{IEEE Journal
  of Selected Topics in Signal Processing}, vol.~9, pp. 1285--1297, 2015.

\bibitem{murguia2020privacy}
C.~Murguia, I.~Shames, F.~Farokhi, and D.~Ne{\v{s}}ic, ``Privacy against state
  estimation: An optimization framework based on the data processing
  inequality,'' \emph{IFAC-PapersOnLine}, vol.~53, no.~2, pp. 7368--7373, 2020.

\bibitem{murguia2021privacy}
C.~Murguia, I.~Shames, F.~Farokhi, D.~Ne{\v{s}}i{\'c}, and H.~V. Poor, ``On
  privacy of dynamical systems: An optimal probabilistic mapping approach,''
  \emph{IEEE Transactions on Information Forensics and Security}, 2021.

\bibitem{Cedric}
E.~Akyol, C.~Langbort, and T.~Basar, ``Privacy constrained information
  processing,'' in \emph{Proceedings of the 54th IEEE Conference on Decision
  and Control (CDC)}, 2015, pp. 4511--4516.

\bibitem{cdc2021arxiv}
H.~Hayati, C.~Murguia, and N.~van~de Wouw, ``Finite horizon privacy of
  stochastic dynamical systems: A synthesis framework for dependent gaussian
  mechanisms,'' \emph{arXiv preprint arXiv:2108.01755}, 2021.

\bibitem{hayati2021finite}
------, ``Finite horizon privacy of stochastic dynamical systems: A synthesis
  framework for gaussian mechanisms,'' in \emph{2021 60th IEEE Conference on
  Decision and Control (CDC)}.\hskip 1em plus 0.5em minus 0.4em\relax IEEE,
  2021, pp. 5607--5613.

\bibitem{prekopa1980logarithmic}
A.~Pr{\'e}kopa, ``Logarithmic concave measures and related topics,'' in
  \emph{Stochastic programming}.\hskip 1em plus 0.5em minus 0.4em\relax
  Citeseer, 1980.

\bibitem{huemer2017component}
M.~Huemer, O.~Lang, and C.~Hofbauer, ``Component-wise conditionally unbiased
  widely linear mmse estimation,'' \emph{Signal Processing}, vol. 133, pp.
  227--239, 2017.

\bibitem{Ross}
M.~Ross, \emph{Introduction to Probability Models, Ninth Edition}.\hskip 1em
  plus 0.5em minus 0.4em\relax Orlando, FL, USA: Academic Press, Inc., 2006.

\bibitem{zhang2006schur}
F.~Zhang, \emph{The Schur complement and its applications}.\hskip 1em plus
  0.5em minus 0.4em\relax Springer Science \& Business Media, 2006, vol.~4.

\bibitem{seber2012linear}
G.~A. Seber and A.~J. Lee, \emph{Linear regression analysis}.\hskip 1em plus
  0.5em minus 0.4em\relax John Wiley \& Sons, 2012, vol. 329.

\bibitem{LogConcave}
M.~Bagnoli and T.~Bergstrom, ``Log-concave probability and its applications,''
  \emph{Economic Theory}, vol.~26, pp. 445 -- 469, 2005.

\bibitem{dharmadhikari1988unimodality}
S.~Dharmadhikari and K.~Joag-Dev, \emph{Unimodality, convexity, and
  applications}.\hskip 1em plus 0.5em minus 0.4em\relax Elsevier, 1988.

\bibitem{hoggar1974chromatic}
S.~Hoggar, ``Chromatic polynomials and logarithmic concavity,'' \emph{Journal
  of Combinatorial Theory, Series B}, vol.~16, no.~3, pp. 248--254, 1974.

\bibitem{johnson2006preservation}
O.~Johnson and C.~Goldschmidt, ``Preservation of log-concavity on summation,''
  \emph{ESAIM: Probability and Statistics}, vol.~10, pp. 206--215, 2006.

\bibitem{marsiglietti2018lower}
A.~Marsiglietti and V.~Kostina, ``A lower bound on the differential entropy of
  log-concave random vectors with applications,'' \emph{Entropy}, vol.~20,
  no.~3, p. 185, 2018.

\bibitem{wellner2012log}
J.~Wellner, ``Log-concave distributions: definitions, properties, and
  consequences,'' \emph{Presentation, University of Paris-Diderot}, 2012.

\bibitem{bobkov2011entropy}
S.~Bobkov and M.~Madiman, ``The entropy per coordinate of a random vector is
  highly constrained under convexity conditions,'' \emph{IEEE Transactions on
  Information Theory}, vol.~57, no.~8, pp. 4940--4954, 2011.

\bibitem{MImatlab}
J.~Delpiano, ``Fast mutual information of two images or signals
  (https://www.mathworks.com/matlabcentral/fileexchange/13289-fast-mutual-information-of-two-images-or-signals),''
  2021.

\end{thebibliography}
\end{document}